# Distributed Adaptive Control: An ideal Cognitive Architecture candidate for managing a robotic recycling plant

Oscar Guerrero Rosado[1,2] and Paul Verschure[1,2,3]

[1] Laboratory of Synthetic Perceptive, Emotive and Cognitive Systems (SPECS), Institute for Bioengineering of Catalonia (IBEC), Barcelona, Spain
[2] Barcelona Institute of Science and Technology (BIST), Barcelona, Spain
[3] Catalan Institution for Research and Advanced Studies (ICREA), Barcelona, Spain
`oguerreros@ibecbarcelona.eu`

**Abstract.** In the past decade, society has experienced notable growth in a variety of technological areas. However, the Fourth Industrial Revolution has not been embraced yet. Industry 4.0 imposes several challenges which include the necessity of new architectural models to tackle the uncertainty that open environments represent to cyber-physical systems (CPS). Waste Electrical and Electronic Equipment (WEEE) recycling plants stand for one of such open environments. Here, CPSs must work harmoniously in a changing environment, interacting with similar and not so similar CPSs, and adaptively collaborating with human workers. In this paper, we support the Distributed Adaptive Control (DAC) theory as a suitable Cognitive Architecture for managing a recycling plant. Specifically, a recursive implementation of DAC (between both single-agent and large-scale levels) is proposed to meet the expected demands of the European Project HR-Recycler. Additionally, with the aim of having a realistic benchmark for future implementations of the recursive DAC, a micro-recycling plant prototype is presented.

## 1    Introduction

Designing a cutting-edge industrial plant in 2020 is a great challenge. Despite the notable growth in a variety of technological areas in the past decade, including Cyber-Physical Systems (CPS), Internet of Things (IoT), cloud computing, embedded systems, Industrial Integration and Industrial Information Integration, the Fourth Industrial Revolution has not been embraced yet.

Strategic initiatives such as Industrie 4.0 (Germany, 2013) and Made-in-China 2025 (China, 2015) represents firm steps toward such a revolution that aims to go further in automation, focusing on end-to-end digitisation and integration of digital industrial ecosystems [1]. Industry 4.0, which is broadly seen as quasi-synonym of Fourth Industrial Revolution, seeks the combination of the following emerging technologies: a) CPS, representing the natural evolution of embedded systems, going from centralised control systems to autonomous machines capable of communicating with each other [2]; b) Cloud computing, that not only provide Industry 4.0 with high-performance computing and low-cost storage but also allow system orchestration by modularisation and sharing resources in a highly distributed way; c) IoT, working as a global network infrastructure that fully integrates identities, attributes and personalities of physical and virtual "Things", thanks to radio-frequency identification and wireless sensor networks [3].

A predecessor of IoT in a country scale industrial context, the Cybersyn Project, can serve us as an example of how important is taking into account the contemporary challenges. This project aimed to collect and transmit economic-related data in real-time to aid Chile's governmental body to make informed decisions in a more democratic manner [4]. Cybersyn began in 1971, however, due to technical, financial and political circumstances met its end in 1973 with Pinochet's dictatorship [5]. To avoid similar failures, any project aiming to get into the Fourth Industrial Revolution must consider contemporary challenges such as: improvement of Information and Communication Technology infrastructures, solving the scalability problem, development of data science and data analytics techniques as well as heterogeneous IoT-related networks. In this work, we address one specific barrier that may hinder progress: the necessity for new architectural models. CPS (e.g. robots) must lead with uncertainty when interacting with the natural world (e.g. industry plant). This uncertainty is due to the changing conditions, the variety of possibilities and the complexity that open environments offer. To tackle this problem, an architecture approach is essential since allows the CPS to be dexterous in different competencies while ensuring safety, security, scalability, and reliability [6]. However, current architectures are not capable of fulfilling all Industry 4.0 requirements. In this paper, a new version of the Distributed Adaptive Control (DAC) architecture is proposed as an ideal candidate to control an industrial plant in a recursive fashion.

Artificial Intelligence approaches have shown promising results when it comes to agents performing simple tasks in dynamic but constrained environments. For example, logarithmic AI solutions have demonstrated successful results (even exceeding human performance levels) in limited domains such as Atari videogames or Go, but implementing a solution that solves a simple navigation task is currently not possible. Furthermore, they require a large amount of training in comparison to human learning [7]. In contrast to board games, industrial plants are highly complex and heterogeneous. Robots operating within such a plant need to be equipped with a wide range of capabilities (i.e. navigation, motor control, human-robot interaction, etc.). Due to the complex behaviour required to these robots, an architectural strategy fits better with the necessities. With an appropriately designed architecture able to organise the different plant-specialised modules and information flow, the system should acquire robustness while performing various tasks. Indeed, the challenge of creating such an architecture opens the question of what design principles must be followed. Although control architectures can accomplish the tasks for which they have been designed, in many cases their success is constrained to a predictable environment, and their performance is far from the human-level efficiency [8]. In contrast, cognitive architectures aim to build human-level artificial intelligence by modelling the human mind. Thus, systems driven by a cognitive architecture could reason about problems across different domains, develop insights, adapt to new situations and reflect on themselves [9].

Here, we propose to create a factory operating system in the form of a synthetic agent, by taking advantage of the fact that a cognitive architecture could work recursively. The term recursive refers to the double-scale functionality of the system: various individual agents that are specialised in different tasks and that are controlled by a higher-level entity, the factory itself. Envisioning a recycling plant with a recursive architecture resonates with the metaphor "Der Mensch als Industriepalast" (Man as an industrial palace) that was proposed by Fritz Kahn. Analogous to a recycling plant, the ingestion of food in Fritz's illustration implies a procedure of treatment, disassembling and classification of the material into different nutrients. For a seamless usage of this disassembled material, the control centre (in the representation of the brain) works on top of the other related systems such as metabolism, blood circulation or respiration.

However, Fritz's illustration depicts a linear process where individual agents are not required to perform dynamically and are hence not involved in the implementation of a cognitive architecture. In contrast, the recycling plant presented here requires a cognitive architecture at both single-agent and large-scale levels, since robots performing tasks of navigation, disassembling, classification, etc. need to work autonomously in a parallel and context-adaptative fashion. The synergic operation of the whole plant depends on the agents' behaviours, which are monitored, controlled and influenced by the large-scale level.

In the following sections, we propose the Distributed Adaptive Control theory of mind and brain (DAC) as a candidate to control a hybrid human-robot recycling plant of Waste Electrical and Electronic Equipment (WEEE) management. Previous work on DAC will elucidate how this architecture supports essential robots' abilities at both single-agent and large-scale levels. Finally, we present a micro-recycling plant as a functional prototype and a benchmark for the implementation of DAC.

## 2      WEEE Recycling Plant

Recycling awareness is gaining importance, especially since recycling plants have to deal with a significant amount of waste per year. The European Union by 2017 recycled and composted 94 Mt (35.2%) of its municipal solid waste [10]. For this reason, and as in many other industries, recycling plants have incorporated a variety of machinery which processes paper, glass, plastics and other materials on a large scale.

However, society is consuming a growing number of electrical and electronic devices that, after a few years, become into e-waste. This e-waste, namely Waste Electrical and Electronic Equipment (WEEE), cannot be processed by the machinery designed to handle the raw materials mentioned previously. The challenge in WEEE management does not only lie in the correct classification of a device but also its disassembly. Each device (such as a TV screen) has a variety of models, and not all models include the same disassembly procedure, or each procedure may be executed in a different order. Additionally, the handling of sensitive or hazardous material (like mercury lamps) adds an extra level of difficulty in the automation of the processing of e-waste. So far, humans seem to be more skilled in performing device disassembly, as their robotic counterparts can only perform partial disassembly that is not generalised to all types of devices. Nonetheless, this partial performance still represents a relief in the arduous task of component disassembling and material. A solution to expedite this work is the development of hybrid human-robot recycling plants where experienced workers would cooperate with specialised robots. This solution implies splitting the disassembling process into subtasks according to the skills of both humans and robots.

A clear example of this new paradigm is the European Project HR-Recycler, where robotic grippers and mobile robots assist in the recycling process of WEEE. Here, robots not only perform repetitive and automated tasks, but they are endowed with autonomous behaviour adaptive to a changing context. Thus, HR-Recycler represents a step towards Industry 4.0 and an ideal framework to introduce new architecture models able to perform the disassembly task even under conditions of uncertainty (like a dynamic, open space in a hybrid human-robot recycling plant). We propose an enhanced plant where we apply the DAC architecture at both single-agent (robots) and large-scale (plant) levels, shaping a multi-scale recursive architecture.

## 3  Distributed Adaptive Control

The Distributed Adaptive Control (DAC) [11] is a theory of the principles underlying Mind, Brain and Body Nexus. It is expressed as a robot-based neural architecture that accounts for the stability maintained by the brain between an embodied agent and its environment through action. DAC assumes that any agent, to act, must continuously solve four fundamental questions, the so-called H4W problem [12]: "Why", reflects the motivation in terms of needs, drives and goals; "What", accounts for the objects in the world that actions pertain to; "Where", represents the location of the object and the self; and "When", serves as a temporal reference of the actions. Additionally, a fifth question (Who) was added, referring to the agency [13].

DAC organises the generation of behaviour horizontally across four layers of control. The Somatic Layer defines the fundamental interface between the embodied agent and its environment, including the needs that must be fulfilled to ensure survival. In a robotic system, this layer accounts for its sensors and actuators and sets its predefined needs. The Reactive Layer provides a set of unconditioned responses working as reflexes for given unconditioned stimuli. This layer represents the first stage of the generation of goals since the behaviours produced follow homeostatic and allostatic principles. The Reactive Layer works on top of the Somatic Layer, gathering sensory data and providing reflex responses through the actuators. A clear example of these reflexes is the "stop signal" triggered when a human gets close to the robot trajectory. The Adaptive Layer frees the system from the restricted reflexive system by perceptual and behavioural learning. It follows classical conditioning principles since the value of the sensory input is shaped by experience, and its outputs could result in anticipation response. Thanks to this layer, the robot can adapt its behaviour according to relevant stimuli (i.e. adjusting its security distance depending on the current scenario). Finally, the Contextual Layer allows the generation of behavioural plans or policies based on sequential memory systems. Sequential representations of states of the environment and the sensory-motor contingencies acquired by the agent are stored in the memory systems, allowing behavioural plan recalling by sensory matching and internal chaining. Thus, the Contextual Layer shows action-dependent learning as observed in operant conditioning, allowing abilities such as allocentric-based trajectory planning, crucial for the mobile robot's navigation. In addition to this layered horizontal organisation, the architecture is also vertically distributed across three columns: states of the world obtained by exosensing, states of the self obtained by endosensing and their interaction through action.

The DAC architecture fulfils both the theoretical and practical criteria that must be considered when designing a proper cognitive architecture. On the one hand, at the theoretical level, DAC applies: a) biologically-inspired learning rules such as Hebbian learning, Oja learning rule or associative competition for different purposes, b) provides a solution to the fundamental Symbol Grounding Problem by acquiring the state space of the agent, based on its interaction with the environment, c) escapes from the now by generating behavioural plans or policies based on sensory matching with representations of environment and action states (stored in memory systems), and d) reverse the Referential Indeterminacy Problem, in which the agent has to extract the external concept that was referred, by endowing the system with proactivity to acquire knowledge. On the other hand, DAC has been validated in both single-agent (i.e. robots) and large-scale levels [14, 15, 16, 17]. These implementations have supported the adequacy of DAC, from a pragmatic point of view, to perform a diverse set of tasks including foraging, object manipulation or Human-Robot Interaction (HRI).

## 4 DAC at the single-agent level

Within the recycling context, we propose two types of robots working as single agents that allow for the transportation, disassembling and classification of the different WEEE components. The two robot categories are robotic grippers and mobile robots. We propose that the implementation of DAC does not differ between robots. However, it needs to be adapted for the different tasks these robots perform, based on the data provided by the different sensors, the needs and goals, and the robots' actuators.

In mobile robots, the needs range in different dimensions depending on their related aCell. Based on the project, we define as aCell the workbench related to a specific worker equipped with a robotic arm where different WEEE is disassembled. Mobile robots aim to transport WEEE in an adaptative way; for instance, taking into consideration the disposal of materials or the specific disassembled and classified components. Here, the Reactive Layer is responsible for driving the needs of the robots towards different navigation patterns as well as pick up and place behaviours, that are carried out by the actuators (i.e. motors of the wheels and the lifting platforms). Sensors such as wheels' encoders, proximity sensors and RGB cameras provide the Reactive Layer with the information needed to trigger reflexive behaviours, and the Adap-tive Layer with the information required to learn associations, which in turn, assist in the behavioural policies formation by the Contextual Layer.

In the robotic grippers, the needs change from navigation-oriented to motor control-oriented goals, since its porpoise is assist in the disassembling procedure itself. Here, the Reactive Layer processes information regarding pressure, proximity and torque sensors, along with data provided by a camera and triggers reflexive behaviours (like a stop signal for safety). This sensory information forms progressively sensory representations by associative learning at the Adaptive Layer and these associations will then be stored in Contextual Layer modules if the associated behaviours allow reaching goal states. Thus, robotic grippers are not just able to perform reflexive actions such as unscrewing. When endowed with adaptive and contextual capabilities, they could, for example, correctly locate the bolt, apply the appropriate velocity and pressure, and predict when the bolt is going to be unscrewed.

At single-agent level, we consider three essential areas for the successful implementation of a recycling plant where robots work alongside humans for e-waste disassembly: Navigation, HRI and Motor Control. DAC has been tested in these areas supporting its implementation in both mobile and humanoid robots.

## 4.1 Previous implementations of DAC: Navigation

Robotic navigation has already been achieved without requiring a cognitive architecture. However, the characteristic of the Industry 4.0 context demands goal-oriented navigation, which adapts to changing environments, needs to be aware of the material transported and the state of the aCell, and ensures the safety of other robots and humans along the trajectory. Due to the complexity of navigation in this context, an architectural approach is more suitable, and the DAC architecture has largely demonstrated its strengths performing foraging tasks with mobile robots.

Aiming to prove that the different computational models proposed by DAC account for functional mapping of specific brain areas and work complementing each other when the system operates as a whole, Maffei et al. [15] embed the version DAC-X in a mobile robot performing a foraging task. In this study, the Somatic Layer computed input signals from the robot's sensors, while the output was calculated as the total motor signal provided by the architecture. Finally, the actions were constrained by the robot's body morphology. The Reactive Layer reflexively mapped sensory states into actions by using feedback controllers that approximated the role of the Brainstem nuclei. Reflexive object avoidance, visual target orientation and computation of bodily states such as needs and drives were obtained by computational models, mimicking the Trigeminal Nucleus, Superior Colliculus and Hypothalamus functions respectively. In the Adaptive Layer, a model of the cerebellar microcircuit allowed associative learning by coupling neutral sensory cues with adaptive responses. The motivation for action arose by modelling the Ventral Tegmental Area for the computation of low-level internal states, and action-selection for behavioural plans was achieved by modelling the Basal Ganglia. Finally, the memory systems of the Contextual Layer comprised a biologically constrained model of the Hippocampus by which the agent acquired an internal representation of the environment; and a model of the Prefrontal Cortex that included mechanisms for storing decision-making and goal-dependent information.

Analysis that discretised the behavioural performance into three phases (early, middle and late trials) showed that at the beginning, the naive agent relies on its reflexes to explore the arena and seek for resources. This initial navigation emerges primarily from the work of the Reactive Layer, resulting in a stochastic trajectory pattern that covered a large part of the arena and had a low item collection rate. After a few trials, the Adaptive Layer took advantage of the local visual landmarks deployed in the floor, thus complementing the primary reactive navigation with adaptive responses. Using the internal representation of space fostered by the Reactive Layer in the explorative trials, the Adaptive Layer allowed goal-directed navigation reducing occupancy of the arena and the mean trajectory length and increasing the collection rate. Finally, at the late trials, the agent ended up displaying a mostly linear trajectory from the home location to the target and back. This linear pattern was achieved thanks to the involvement of the Contextual Layer since it combined a robust representation of the environment and made available the goal locations stored in the long-term memory. These achievements in a hoarding task, by implementing DAC in a mobile robot, support a similar implementation in the context of Industry 4.0.

## 4.2    Previous implementations of DAC: Human-Robot Interaction

In the context of a recycling plant, social skills such as empathy, natural language, or social bond formation do not precisely fit the context of collaboration between humans and robots in an industrial plant. However, the human workers will interact with the robots, especially the robotic grippers, as they both will be required to perform tasks with the common goal of disassembling a device. These robots will also consider that workers may show different skills, preferences, and even trust in robots. Perceived safety, collaboration, adaptation to each worker and the completion of a task are critical points for successful Human-Robot Interaction.

An example of a successful collaboration that ensures safety has been shown by [18]. Here, the authors presented a collaborative human-robot assembling task. However, the task was restricted to the assembling of a single component, always following the same steps, and no tool manipulation was required. In contrast, the disassembling process requires tool manipulation and includes a variety of devices, where the disassembly steps may differ. Thus, a more adaptive solution is needed to disassemble different WEEEs while collaborating with the human partner successfully.

Numerous contributions to the field of HRI have been provided by DAC through its implementation in humanoid robots. However, focusing on relevant problems for recycling plants, addressing the anchoring problem is essential. The anchoring problem refers to the process of creating and maintaining the link between raw data provided by the sensors and symbolic representation processed and stored by the system. [16] tackled this problem by defining a representation of knowledge based on the so-called H5W problem. Thus, several entities connected by semantic links (who, how, what, where and when) full describe the situation. In other words, the Somatic Layer is taking the sensory input data (i.e. spatial properties of objects and agents). Subse-quently, the Adaptive Layer will translate this data into instances (i.e. unsafe situation) by providing solutions to the H5W problem. Finally, these solutions will be compared with those stored in the Long-Term Memory of the Contextual Layer (that previously showed good results), so the best one will be selected.

In the context of a recycling plant, considering the limited social skills of a robotic gripper, and at the same time that interaction with the worker is needed, we propose two channels of communication. As most plants are considered noisy, and workers wear protection gear, verbal communication is not preferable. For this reason, we will employ a predefined gesture-based communication of a set of fundamental requests to the robot, such as start, stop, take rest position, etc., by providing the workbench with a computer vision system able to solve the anchoring problem. Gestures, combined with a multitouch interactive tablet, will provide more complex interaction scenarios.

### 4.3 Previous implementations of DAC: Motor Control

Simple behaviours such as grabbing a tool, unscrewing a bolt or extracting and placing a component during the disassembling task are complex movements that require lengthy training sessions until a robotic gripper can adequately perform such actions.

Although a complete version of DAC has not been applied to control the specific behaviours of a robotic gripper, previous studies have validated the implementation of the DAC architecture for motor control. More specifically, motor control was achieved with the acquisition of affordances, namely the categorisation of goal-relevant properties of objects [19]. Within the context of DAC, Sanchez-Fibla, Duff & Verschure [20] proposed the notion of affordance gradients: object-centred representations that describe the consequences that an action may have on this particular object. Through object-centred force fields, the agent was not just able to predict the outcomes of an action, but also to generalise predictions to actions that the agent has not previously perform. These affordance gradients were acquired through learning in Adaptive Layer, allowing to a mobile robot to push an object from the right side and place it in a target position and orientation. These affordance gradients were recently extended to the acquisition of bimanual affordances in Sanchez-Fibla et al. [21].

## 5 DAC at the large-scale level

Interestingly, DAC has also shown its capabilities to control an entertainment space. Ada [17] was a large-scale intelligent and interactive environment that was able not just to learn information from its visitors, but also to modify its behaviour guiding their steps toward a given direction. Ada achieved interaction with its visitors by expressing its internal states through global lighting and background sound. Information processing through DAC allowed leveraging multi-modal data from Ada's sensors (cameras, microphones and pressure-sensitive floor) to learn the best way to interact with its visitors following paradigms of classical and operant condition, demonstrating that DAC is not constrained to conventional robots.

## 6      Multi-scale DAC: A micro-recycling plant

For a recursive multi-scale implementation of DAC architecture, a central system is endowed with DAC operating on a large-scale level, and so controlling the synergic functioning of the single-agent level (Fig.1). DAC at this large-scale level is implemented more abstractly, since it leverages sensors and effectors of the single-agent level, leading to an intertwined recursive multi-scale architecture. The fact that this central system integrates information from all single agents allows new perceptions such as the amount of aCell that are free or taken, space occupied by mobile robots or mean amount of WEEE disassembled.

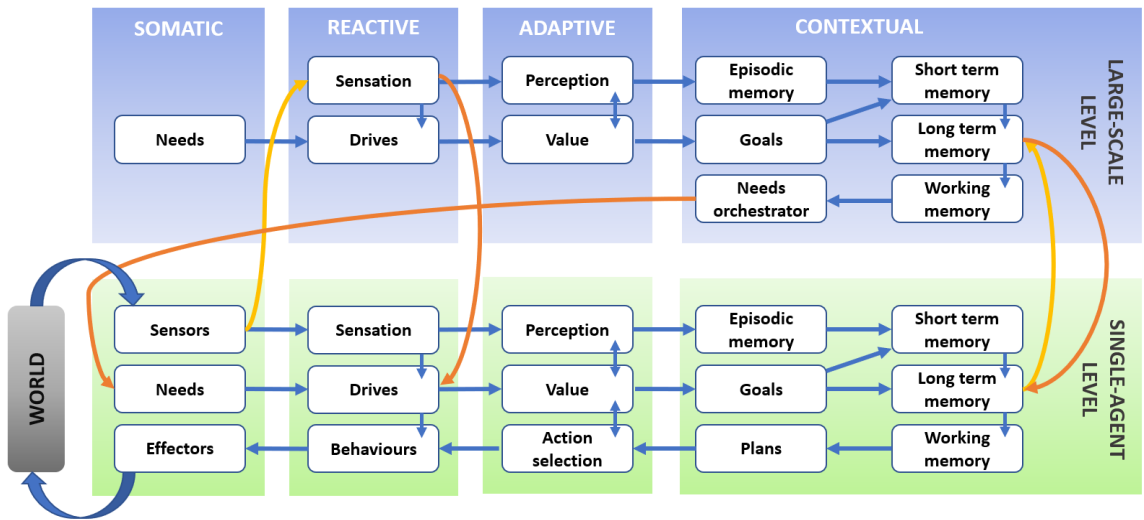

**Fig. 1. Recursive DAC.** Arrows represent information flow. Blue arrows indicate a connection between modules and layers of the same entity; Yellow arrows represent information sent from single-agent level to large-scale level; Orange arrows represent information sent from a large-scale level to single-agent level.

By using a wireless connection, the large-scale level creates a network with each single-agent, consisting of three loops. A sensory loop integrates data from the different robots' sensors at the large-scale level, allowing overall interpretation of the context and therefore triggering reflexive signals to every robot (e.g. stopping signals in case of general danger situation). Based on the needs of the large-scale level and the current state of the plant, an orchestrator loop is in charge of modulating the needs of every single agent. This second loop allows the robot to behave in an allostatic way between worker-based and plant-based needs. A third loop is in charge of interconnecting Long-Term Memory modules across the entire plant. By connecting the LTM module of the large-scale level to those LTM modules embedded in each robot, learning generalisation and information sharing occur across single agents. Thus, workers could find the aCell adapted to their needs even if they change from one workbench to another, or mobile robots could plan their trajectories based on the location and trajectories of others.

To evaluate the candidature of DAC as a perfect candidate architecture to control an industrial plant within the context of Industry 4.0, we are developing a prototype of a micro-recycling plant.

### 6.1 Micro-plant design.

To build the closest setup to the HR-Recycler project, our design for a robotic micro-plant includes both mobile robots and robotic arms (Fig.2). These robots also embed those sensors used in the project (cameras RBG, proximity and pressure sensors, wheels' encoders, etc.). In this prototype, workers are represented by balancing robots that, by using a visual cue, are related to a specific workbench and embed different worker's characteristics. However, unlike HR-Recycler, these robotics workers will no assist in the disassembling process. The robotics arms will be in charge of full disassemble simple devices composed of four parts representing different materials (plastic, metal, paper and risky material). To classify these components, robotics arms also will place the parts of the components into bins coloured according to the material that must be collected in it. Using a lifting platform, mobile robots can lift the bins and transport it when is full. A computer placed outside the micro-plant will be running the central control system that allows synergic performance between all the agents implicated in the recycling plant. Additionally, a conveyor belt will be used to facilitate intermediate steps in the development process.

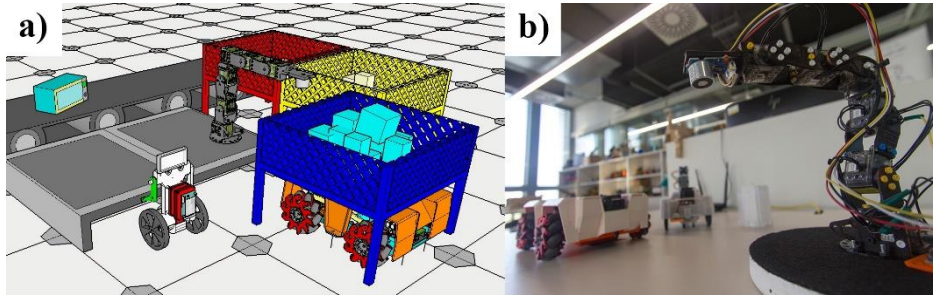

**Fig. 2. Micro-recycling plant model and robots.** a) 3D sketch of single-agents working synergistically to disassemble WEEE. Mobile robots can lift and transport coloured bins. These bins are coloured according to the material that must be classified in it. Robotics arms adapt its performance based on the presence of a worker represented by balancing robots moving around the plant. b) Functional robots to be implemented in the micro-plant.

### 6.2 Future benchmarks and expected results.

After testing DAC architecture runs correctly in each of the agents individually, we propose two benchmarks in order to assess the success of the recursive architecture proposed. First, we will evaluate the implemented multi-agent navigation systematically deploying two or more mobile robots that will operate under two conditions:

autonomously with the large-scale influence of the plant, and autonomously but without this influence. When the central control system is not enabled, we expect to see navigation adaptative to the contextual characteristics of the environment and goal-oriented behaviour related to the transportation of material from or towards the workbenches. However, coordination between the agents will be not found, unless the central control is enabled, leading to convergence of trajectories and no distribution of spaces, workbenches and materials. Second, we will evaluate the generalisation of worker characteristics by deploying two different robotics arms performing in an adaptative way to the worker situated in its related workbench. With the central control system not enabled, we expect to find adaptive behaviours of both robotics grippers toward their related worker (i.e. distance to the worker based on its trust in robots). However, if the workers are exchanged of the workbench, the adaptative behaviour to the specific worker performer by the previous gripper will not be found in the new place, unless the central control system is enabled.

## 7    Discussion.

In this article, we proposed the Distributed Adaptive Control (DAC) cognitive architecture as a candidate for robot and plant control within the context of Industry 4.0. The implementation of this architecture has been supported by previous works on different robotic areas such as Navigation, Human-Robot Interaction and Motor Control. However, the following issues have not been discussed yet.

The implementation of DAC has been addressed within the context of a hybrid human-robot recycling plant. This kind industrial plant can be perceived as a simplified instance since in comparison with the overall idea of Industry 4.0 it less dependent on other technologies such as IoT or cloud computing. However, although we are convinced that our recursive implementation of DAC could take great advantage of such technology, the Fourth Industrial Revolution will be achieved thanks to discreet but firm steps.

How much must be learned and how much must be prewired by the agents is another important question to solve, in order to maintain both adaptability and efficiency. We propose that basic abilities such as grabbing a tool or creating and navigating a map of the environment are preferably achieved in previous training sessions, so the agent can adapt a behaviour already learned. Hence, the robots just should adapt these abilities already learned to the position of the tool or the trajectory of other mobile robots. Other significant information such as the referred to the worker's abilities and preferences could be integrated directly in the central control system by using questionnaires.

**Acknowledgements.** This material is based upon work funded by the European Commission's Horizon 2020 HR-Recycler project (HR-Recycler-820742H2020-NMBP-FOF-2018).